\documentclass[runningheads]{llncs}
\usepackage[T1]{fontenc}
\usepackage{graphicx}
\usepackage[colorlinks=true,linkcolor=blue,citecolor=blue,urlcolor=blue]{hyperref}

\begin{document}
\title{Overview of Current Challenges in Multi-Architecture Software Engineering and a Vision for the Future}

\titlerunning{Current Challenges in Multi-Architecture Software Engineering}

\author{Piotr Sowiński\inst{1,2}\orcidID{0000-0002-2543-9461} \and
Ignacio Lacalle\inst{3}\orcidID{0000-0002-6002-4050} \and
Rafael Vaño\inst{3}\orcidID{0000-0003-2372-6253} \and
Carlos E. Palau\inst{3}\orcidID{0000-0002-3795-5404} \and
Maria Ganzha\inst{1,2}\orcidID{0000-0001-7714-4844} \and
Marcin Paprzycki\inst{2}\orcidID{0000-0002-8069-2152}
}
\authorrunning{P. Sowiński et al.}
%
\institute{Faculty of Mathematics and Information Science, Warsaw University of Technology, ul. Koszykowa 75, 00-662 Warsaw, Poland \and
Systems Research Institute, Polish Academy of Sciences, ul. Newelska 6, 01-447 Warsaw, Poland\\
\email{\{piotr.sowinski,maria.ganzha,marcin.paprzycki\}@ibspan.waw.pl} \and
Communications Department, Universitat Politècnica de València, 46022 Valencia, Spain\\
\email{\{iglaub,ravagar2\}@upv.es, cpalau@dcom.upv.es}}
\maketitle              
\begin{abstract}
%
The landscape of computing technologies is changing rapidly, straining existing software engineering practices and tools. The growing need to produce and maintain increasingly complex multi-architecture applications makes it crucial to effectively accelerate and automate software engineering processes. At the same time, artificial intelligence (AI) tools are expected to work hand-in-hand with human developers. Therefore, it becomes critical to model the software accurately, so that the AI and humans can share a common understanding of the problem. In this contribution, firstly, an in-depth overview of these interconnected challenges faced by modern software engineering is presented. Secondly, to tackle them, a novel architecture based on the emerging WebAssembly technology and the latest advancements in neuro-symbolic AI, autonomy, and knowledge graphs is proposed. The presented system architecture is based on the concept of dynamic, knowledge graph-based WebAssembly Twins, which model the software throughout all stages of its lifecycle. The resulting systems are to possess advanced autonomous capabilities, with full transparency and controllability by the end user. The concept takes a leap beyond the current software engineering approaches, addressing some of the most urgent issues in the field. Finally, the efforts towards realizing the proposed approach as well as future research directions are summarized.

\end{abstract}

\section{Introduction}

The modern landscape of hardware and software platforms is changing rapidly. In CPU architectures, the existing technologies are expanding their ranges of applications (e.g., ARM processors shifting to the servers), while entirely new platforms, such as RISC-V~\cite{greengard2020will}, emerge. This requires the existing software to be adapted to the new instruction sets, which can be a costly and lengthy process. At the same time, computational workloads are increasingly deployed across the entire computing continuum~\cite{gkonis2023survey}, introducing even more hardware diversity. Software engineers are faced with the seemingly impossible task of producing applications that are required to run on multiple highly heterogeneous platforms, spanning from the large Cloud clusters to the tiny IoT devices. The software must also evolve rapidly, to keep up with consumer demands, technological changes, and to maintain competitiveness in the global market. 
In this context, the current specific challenges for software vendors include migrating Cloud applications to the Edge~\cite{vano2023cloud}, and using the emerging, quickly evolving hardware (e.g., specialized machine learning accelerators~\cite{alam2024survey,peccerillo2022survey}) to its full potential.

Much of the recent innovation in software engineering (SE) has been focused on increasing the degree of automation in all stages of the software lifecycle -- from the Continuous Integration / Continuous Delivery (CI/CD) pipelines~\cite{soares2022effects} to the artificial intelligence (AI) based coding assistants, and documentation generation tools~\cite{hou2023large}. However, the current automation tools are, generally, disconnected from each other. They lack autonomy in making decisions, and are usually regarded (and designed) as corrective patches that aim only to heal a specific issue, thus lacking a holistic view of the software lifecycle and the variety of existing interdependencies. Effectively, the whole SE process still hinges on the developer being able to grasp the full complexity of the software, and initiating all required actions. Considering that, introducing autonomy to the SE tools could present a deep transformative change in how it is possible to tackle complex problems, such as resource use optimization of multi-architecture software, or adapting applications to the dynamically changing environment.

Independently, there is also a clear need to improve software modeling capabilities, moving beyond the isolated models that address only a narrow slice of the software lifecycle. Integrated, yet highly modular and dynamic models are needed. They must be comprehensible to both humans and computers, forming a much-needed bridge of mutual understanding, allowing both sides to operate with the same concepts (i.e., AI alignment). 
Such holistic models could become the basis of future software engineering automation tools that could carry out complex tasks autonomously, while remaining completely transparent, explainable, and controllable.

\paragraph{Guiding example.} Throughout the text, to better illustrate the presented concepts, we will be referring to a guiding example, based on a real use case from the Smart Safety of Workers pilot in the ASSIST-IoT Horizon 2020 project~\cite{danilenka2023real}. This scenario is meant to be challenging from the software engineering standpoint, highlighting areas in need of improvement. 

In the considered scenario, a software module that can detect falls of construction workers in near-real time is to be developed. The module uses acceleration measurements collected from IoT devices worn by the employees. It must work on a very wide variety of hardware platforms, depending on the client and the existing computing infrastructure. Here, it is assumed that different construction companies may have different (e.g. legacy, present or future) hardware and software platforms. This includes highly heterogeneous IoT and Edge devices (used for lower latency and better privacy), as well as Cloud clusters, e.g., in the case of clients who do not have enough computing capacity on the Edge. Moreover, it is assumed that the used hardware can change quickly (even every few months) due to the shifts in the device availability, client preferences, technological advances, and older devices breaking quickly in the harsh environment. The hardware platforms may have different processor architectures (including x86, ARM, RISC-V), computing capabilities (from hundreds of kilobytes to terabytes of system memory), machine learning accelerators, and networking (Ultra-Wide Band, Wi-Fi, Bluetooth, ZigBee...). The worker safety application must be able to evolve very quickly to, among others, adapt to client demands and/or newly-introduced machine learning models, and to keep up with the competition. For example, new integrations with clients' systems are expected to be one of the most common requests. The clients also want to understand the system's design and evolution, to make sure it reflects their requirements in terms of, among others, use case requirements (e.g., workplace safety standards), cybersecurity, or worker privacy.

\paragraph{Summary of contributions.} In this work, we consider the intersection of the following three relevant challenges in modern software engineering: \textbf{(i)} developing multi-architecture applications, \textbf{(ii)} accelerating and automating the SE processes, and \textbf{(iii)} software modeling. For each, a deep dive into the current technological obstacles and opportunities is presented. We believe that these challenges are interconnected and thus propose a joint solution -- an architecture of a general system accelerating software engineering tasks for multi-architecture applications using WebAssembly. The presented system concept targets all stages of the software lifecycle -- from planning, through coding, testing, up to deployment and operation. It leverages the latest achievements in AI, autonomy, and knowledge representation to potentially pose a paradigm shift in software engineering and change how next-generation software is developed.

\section{Background}\label{sec:background}
In what follows, we present an analysis of the main challenges and latest developments in the three considered areas: multi-architecture applications, software engineering process acceleration and automation, and software modeling.

\subsection{Multi-Architecture Applications}\label{sec:background:wasm}

Multiple challenges in building software for multi-architecture systems reflect the complexity and the heterogeneity of modern distributed computing environments, primarily in the computing continuum~\cite{kumar2021coding,sowinski2023autonomous}. The used platforms span from the large compute clusters (e.g., in the Cloud), to the resource-constrained IoT sensors and actuators. This being the case, the developers must be aware of all possible CPU Instruction Set Architectures (ISAs) present in the devices, and have dedicated tooling for each. The hardware platforms also differ greatly in terms of supported compute accelerators, storage, and networking capabilities. Meanwhile, the need to make the applications more dynamic and resilient is undeniable -- software providers must cope with high availability requirements, rapidly changing customer demands, diverse hardware, and supply chain risks. For example, in the event of a chip shortage, the ability to effortlessly switch to a different ISA, for which processors are more widely available, would be of tremendous value to the businesses. Although it is possible to tackle these issues partially with the current tools (as described in what follows), it is generally a difficult task, requiring expertise not only in coding, but also in compilers, distributed deployments, self-adaptation, and more, which increases the costs.

Currently, partial solutions to these problems exist, and are in active use in the industry. Containers are perhaps the most popular~\cite{vano2023cloud,casalicchio2019container}, providing a semi-isolated runtime environment and some degree of portability. However, their multi-architecture capabilities rely on building the container image separately for each ISA, which requires all dependencies to also be available for that instruction set. In the past, container developers usually prepared only images for x86 or x86-64 processors. Through concerted effort, the industry in the recent years managed to make ARM-based container images also widespread~\cite{kaiser2022container}. However, the limitations of this approach become visible with the emerging RISC-V architecture~\cite{dakic2024evaluating}. If RISC-V is to see wide adoption, every software vendor will have to again adapt their software to the new ISA. This recurring adaptation process slows down the take-up of new hardware platforms, and substantially increases the costs of multi-architecture application development. As an alternative to the containers, one may consider programming languages that compile to a portable bytecode, which can be then executed on any supported hardware platform. Such is the case with the Java Virtual Machine (JVM), or .NET. However, both of these frameworks support only a limited number of languages which were specifically designed or ported to compile to the bytecode (e.g., Java for the JVM, C\# for .NET).

When considering the guiding example, the presented technologies offer no satisfactory solutions. The company developing the worker fall detection application would most likely choose to maintain a few versions of the software in parallel. For example, for the larger Cloud and Edge devices, multi-architecture containers would be used, with each ISA being supported separately. For IoT devices, the application would have to be recompiled into a binary format specific to a given platform, making the process more expensive. Additional consideration would have to be given to the machine learning accelerators, available on each platform, as they often require specialized libraries or runtimes to be used, further increasing the amount of the needed binary distributions. For the clients requiring additional integrations, separate distributions including additional code would also be needed.

One of the most promising recent developments in this area is WebAssembly (Wasm)~\cite{Rossberg:24:WCS} -- an emerging technology for building multi-architecture executables. Although it was initially designed to boost the performance of browser-based applications, it is increasingly being used in general-purpose systems~\cite{vano2023cloud}. Its advantage over previous approaches (e.g., the Java Virtual Machine) lies in: \textbf{(i)} a very diverse set of supported programming languages, \textbf{(ii)} a much lighter and modular design, \textbf{(iii)} support for a wider variety of hardware platforms (including the smallest IoT devices~\cite{ray2023overview}), and \textbf{(iv)} the involvement of many large and competing technological stakeholders from the start, instead of a single, dominating entity. Specifically, some of the most significant members of the Bytecode Alliance (the nonprofit organization maintaining parts of the Wasm ecosystem) include Intel, Amazon, ARM, Cisco, Microsoft, Mozilla, Red Hat, and Siemens\footnote{\url{https://bytecodealliance.org/}}. With a much smaller footprint and better security than containers~\cite{vano2023cloud,menetrey2021twine}, Wasm appears to be the execution model of the future computing continuum applications~\cite{menetrey2022webassembly,sowinski2023autonomous}. Its key advantage is the ability to run a WebAssembly binary on almost any platform in the computing continuum.

Although Wasm is a relatively new technology, its ecosystem is developing rapidly. Among key initiatives in this space is the WebAssembly System Interface (WASI)\footnote{\url{https://wasi.dev/}}, which seeks to provide a portable and secure system interface~\cite{spies2021evaluation}, including functionalities such as sockets, filesystem access, and the HTTP protocol. The WASI initiative is currently evaluating proposals for even more ambitious features, such as uniform interfaces for machine learning acceleration, GPU access, SQL databases, and key-value stores. In this sense, WASI may grow beyond what was traditionally thought of as a ``system interface'' (e.g., POSIX~\cite{lewine1991posix}), by providing a universal interface for a wide variety of common functionalities that may be used across the computing continuum. At the time of writing (October 2024), the latest release of WASI is 0.2.2, released on October 4, 2024. The next feature version (0.3) is expected to include the introduction of asynchronous APIs.

The Wasm Component Model\footnote{\url{https://component-model.bytecodealliance.org/introduction.html}} is a recent architectural proposal, aiming to enable component composability in WebAssembly binaries. With the Component Model it is possible to seamlessly link together modules written in a variety of languages, which are normally not easily interoperable -- for example C++, C\#, Go, JavaScript, Python, and Rust. The proposal includes a universal binary interface language (Wasm Interface Type, WIT) in which contracts between components are defined. This is an ambitious project, as each of these languages has a completely different and unrelated type system. This is in contrast to, for example, the Java Virtual Machine, in which all languages share the same basic JVM constructs and types. Concurrently with the Wasm Component Model, a registry protocol for Wasm components is being developed: warg\footnote{\url{https://warg.io/}}. The warg protocol envisions a federated network of Wasm registries which will engage in, for example, collaborative vulnerability tracking.

Finally, wasmCloud\footnote{\url{https://wasmcloud.com/}} should be mentioned, as an orchestration framework for WebAssembly applications, explicitly targeting heterogeneous deployments and Cloud-Edge scenarios. It builds on top of the component model and WASI, to make application deployment as seamless as possible. Additionally, wasmCloud provides a mesh network (called ``lattice'') based on NATS\footnote{\url{https://nats.io/}} to enable communication between components.

Despite these recent advances, it should be stressed that at this time, the WebAssembly ecosystem has not yet fully matured. The aforementioned projects are all working towards a new common technological baseline that is expected to crystallize in the next few years. Developers migrating to Wasm are encountering practical issues in their code that take time to resolve~\cite{waseem2024issues}. WebAssembly is also still lacking in overarching software engineering methodologies and tooling, to create a robust, complete software lifecycle. Such methodologies and tools are needed, given the radical change in the underlying technologies and assumptions, which question the applicability of past approaches.

In the guiding example, it can be stipulated that the company would use Wasm to build a universal module that can work on any platform. Interfacing with the network and the machine learning accelerators would be done via WASI. Client-specific code would be implemented and composed with the Wasm Component Model. The complete deployment would be then orchestrated via wasmCloud. However, WebAssembly by itself does not solve all challenges. Additionally, a new SE methodology and toolset would be needed to adapt the company's processes. Let us look into these issues in more detail in the following subsections.

\subsection{Software Engineering Process Acceleration and Automation}

Accelerating software development and maintenance has been the focus of software engineering for decades~\cite{raccoon1997fifty,osterweil2008determining}. Although much was achieved in terms of methodologies, tools, and industry practices, the constantly increasing complexity of software processes still requires more innovation. Inefficient processes can be found across the entire software lifecycle, with frequent miscommunication, lack of traceability, and monolithic changes that are hard to manage. For example, in requirements analysis, miscommunication between the stakeholders and the technical teams is common. In the code writing phase, the teams often lack precise information on what to implement and what the user really needs. Meanwhile, in code maintenance, it is hard to avoid regressions due to partial or poorly understood bug fixes, or cascading dependencies that make on-the-fly modifications very challenging (if not impossible) to automate.

Effectively automating processes in software development and maintenance is crucial to ensuring timeliness, high quality, and consistency of the software product. Automation can reduce costs and greatly accelerate software development~\cite{soares2022effects}. However, the automation capabilities of the state-of-the-art tools are still limited, and do not always deliver on the expectations~\cite{fitzgerald2014continuous,wiklund2017impediments}. As a result, many processes still rely on the developers to carry out the critical work manually. Although maintaining control of the processes is crucial due to (among others) security and quality assurance concerns, this does not preclude extending automation to making autonomous decisions. Controllable, transparent, and explainable autonomy of software workflows would enable one to realize use cases that were previously very hard or even impossible to implement reliably. This includes program self-repair, run-time self-adaptation, and more, taking over the decision-making burden from the developer. Micro-managing the complexity of modern software appears to be no longer possible to be realized by hand. In the long term, continuing improvements in resource efficiency, security, and privacy may only be achievable with better automation.

In this context, the last two years have seen perhaps the most abrupt paradigm shift in software engineering to date. The emergence of Large Language Models (LLMs) opened up new possibilities for automation, such as coding assistants (e.g., GitHub Copilot~\cite{dakhel2023github}), documentation generation (e.g., Scribe, DocuWriter), testing, and more~\cite{hou2023large}. However, LLMs have obvious issues with transparency and reliability~\cite{bender2021dangers,weidinger2022taxonomy}, which has serious practical implications for SE~\cite{sallou2024breaking}. For example, language models for code generation were demonstrated to be prone to misusing APIs and introducing new vulnerabilities to the applications~\cite{mousavi2024investigation}. LLMs are also still largely disconnected from the SE methodologies and from other existing tools. Modern software processes are already very complex, with a multiplicity of tools that assist developers during each stage of the development lifecycle. Introducing AI into the mix in an \emph{ad hoc} manner further complicates the situation. This poses the question of whether further optimizations in software processes should instead be sought in a more holistic and systematic manner.

Although process automation is widespread in modern software engineering, the capabilities that are delivered by state-of-the-art tools remain limited. Numerous DevOps and DevSecOps~\cite{lopez2021devsecops} platforms are commercially available (e.g., GitLab, GitHub), along with many CI/CD automation tools, either integrated into such platforms or external (e.g., Jenkins, CircleCI, Travis). However, these tools must be set up manually, thus requiring deep expertise in CI/CD tooling~\cite{rajapakse2022challenges}. Furthermore, such tools have no or very limited autonomous capabilities -- usually they operate on a set of fixed triggers, also set up manually. On the other hand, with the recent surge in the popularity of LLMs, AI-based coding assistants (e.g., GitHub Copilot) offer to automate some tasks, such as code writing, commit message management, or writing task descriptions. However, these tools also require explicit prompting from the user and take no action on their own -- they do help accelerate certain activities, but possess no autonomous decision-making capabilities. While early work was recently done in building autonomous coding agents~\cite{wang2024survey}, they rely heavily on black-box, unaccountable LLMs, and very little is known about their effectiveness in practice. However, it is important to note that as of today (October 2024), research concerning relations between LMMs and SE is one of the hottest research areas.

In the guiding example, the fall detection application will get more complex over time, making it increasingly difficult to maintain. For each new platform, introducing additional CI/CD configuration will be required. It would also be very challenging (or impossible) to automate adaptations to changing client requirements, or to automate security vulnerability patching (which is one of the most important forms of necessary software maintenance).

\subsection{Software Modeling}

The increasing complexity and dynamicity of modern software pose a major challenge for software vendors. Nowadays, applications are composed of hundreds of software modules and libraries. As such, modeling the intricate web of implemented features, requirements, code, and dependencies is a slow and laborious task. At the same time, the application as a whole must evolve rapidly, to make sure that the ever-shifting demands of the stakeholders are met, and to stay relevant in the very competitive market. Not being able to efficiently manage this growing complexity can lead to wasted resources (e.g., developer time, electricity), security issues, bloated applications, or delays in integrating software components. Conversely, an accurate, representative model of the application and its requirements could drastically accelerate development and maintenance, and serve as a reliable ground truth for both the developers and the automated tools.

The increasing complexity of software projects, and the need to consider the entire lifecycle of an application strain the existing Model-Driven Engineering approaches~\cite{bucchiarone2020grand}. Current solutions must be made more dynamic, scalable, and enable intelligent automated completion in the model specification and creation. Although UML is the industry standard in software modeling, using it for communication with stakeholders and among team members is time-consuming and prone to misunderstandings~\cite{fernandez2018industrial}. BPMN is more business- and process-oriented, with some overlap with UML. However, in practice, modeling errors are common in BPMN models~\cite{leopold2015learning}. Moreover, it is considered overcomplicated and hard to understand by some users~\cite{recker2010opportunities}. Moving on to the code, design-by-contract programming formally specifies the behavior of software components~\cite{meyer1992applying}. However, this is usually only applicable within a single programming language, and is disconnected from other types of software-related models. Although there are some solutions for model-driven code generation~\cite{sunitha2019automatic} and other AI-based tasks, especially using LLMs~\cite{camara2023assessment}, the range of their capabilities is very narrow, and it is not clear how to integrate them with each other, and with the existing SE processes. Moreover, the aforementioned modeling approaches only consider a part of the software lifecycle (e.g., requirements, code, dependencies) and can hardly be extended to the other application areas. 

The problem of having many disconnected solutions to software modeling could be addressed with an approach that is modular, expressive, and freely extensible from the start, such as ontologies and knowledge graphs~\cite{Grimm2011,hogan2021knowledge}. Modern ontology languages like OWL~\cite{OWL} use the Semantic Web as their technological backbone, which gives them the ability to interlink knowledge across the Internet~\cite{bizer2008linked}. Indeed, several ontologies for software engineering applications were already proposed. For example, the Software Build System Ontology (SBSON)~\cite{eghan2019dependency} allows for modeling dependencies, releases, and software packages. On the other hand, the TwoUse Toolkit~\cite{parreiras2012semantic} translates UML class diagrams into ontologies. The Software Description Ontology~\cite{garijo2021software} provides terms for describing software, especially applications developed for scientific purposes. Unfortunately, none of these approaches resolves the software modeling issue completely and holistically. However, they are not required to do so. They only need to cover well the part of the software lifecycle relevant to them. It is possible to later reuse multiple pertinent ontologies in concert, to obtain a holistic solution tailored to a given software engineering use case. This, however, is still an unexplored area of research.

The ontology-based approach allows one to interlink knowledge not only within one domain (software engineering), but also beyond it, opening new possibilities. Depending on the use case, it could be for example beneficial to interlink the software model with a description of research artifacts (e.g., benchmark results) as provided by the Information Research Artifact Ontology~\cite{nguyen2021ontology}. When considering the deployment of distributed sensor networks and similar systems, one may use standard ontologies such as Web of Things TD~\cite{Korkan:23:WTT}, SOSA/SSN~\cite{Phuoc:17:SSN}, or SAREF~\cite{garcia2023etsi}. Finally, especially relevant in requirements modeling and communicating with the stakeholders, accurately representing the specific application domain is often crucial. Ontologies can also help with that, as they were successfully applied across a wide range of use cases, from biomedicine~\cite{konopka2015biomedical} to building information modeling~\cite{niknam2017shared} and popular music~\cite{weigl2021fair}. This Semantic Web-based approach breaks with the usual assumptions of what a software model is and, essentially, puts no boundaries or limits on what can be modeled in a uniform manner. This, however, requires a lot of work to be put into adhering to the Linked Data best practices and the FAIR principles (findability, accessibility, interoperability, reusability)~\cite{poveda2020coming}. Otherwise, practical issues in reusing ontologies may create too much friction for adopters and make the proposed vision of an ecosystem for knowledge graph-based software modeling rather unattractive.

In the guiding example, the company would most likely use UML (or BMPN, or a similar tool) to communicate with the clients. However, maintaining the models for each client and each software variant would be very time-consuming, and the clients may not be familiar with the selected modeling tools. Alternatively, the company could employ knowledge graphs to represent the use case logic and requirements, however, translating this knowledge into the actual software is an open research subject. The knowledge graph would also have to be constructed and then actively maintained, which in practice can turn out to be a serious problem, involving long-term resource commitment~\cite{zablith2015ontology,zhong2023comprehensive}. Therefore, for this approach to succeed, the process of knowledge elicitation and formalization, as well as management of knowledge evolution would have to be at least partially automated. This, again, is an area of active research. 

\section{Proposed System Architecture}

\begin{figure}[htb]
    \centering
    \includegraphics[width=\linewidth]{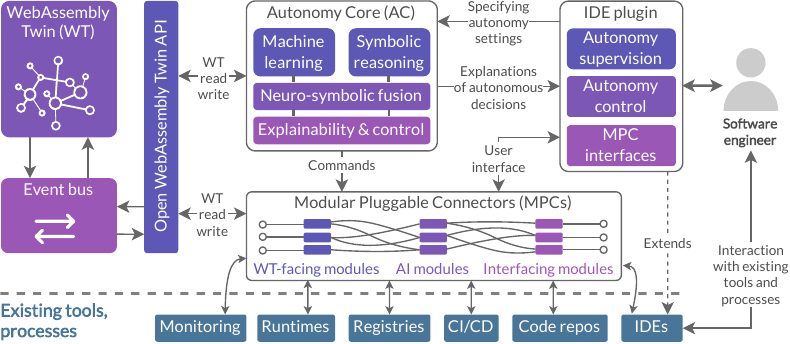}
    \caption{Overview of the proposed system architecture.} 
    \label{fig:overview}
\end{figure}

Taking into account what has been discussed thus far, let us introduce an architecture for a system that would tackle the indicated challenges of software modeling and SE process automation for multi-architecture software. The proposed concept offers a comprehensive view of the entire software engineering lifecycle, and can plug in seamlessly into the existing tools and processes, such as those discussed in Section~\ref{sec:background}. The general architecture is described in what follows, with a visual overview presented in Figure~\ref{fig:overview}. Further details about each component (WebAssembly Twin, Autonomy Core, and Modular Pluggable Connectors) are presented in subsequent sections.

The technological and conceptual backbone of the proposed architecture is the \textit{WebAssembly Twin} (WT), a dynamic, knowledge graph-enabled model of the WebAssembly software. The WT is intended to serve as a closely tied, bi-directionally synchronized ``twin'' of the WebAssembly modules and applications. Not only does the WT model reflect the actual, existing software, but also any change in the model induces corresponding changes in the software -- essentially, providing a powerful Model-Driven Engineering abstraction. The WT is inherently modular, with modules corresponding to the different versions of the software and to the different modeling domains. The WT uses dynamic knowledge graphs~\cite{polleres2023does} with event-driven reads and writes, flowing through a high-performance event bus. Access to the WebAssembly Twin is provided via an open, universal API that is used by the Autonomy Core (AC) and the Modular Pluggable Connectors (MPCs). The WebAssembly Twin is further discussed in Section~\ref{sec:arch:wt}.

\textit{Modular Pluggable Connectors} are non-autonomous automation tools -- executing specific commands, to carry out some work in the software (via the existing tools, processes, and platforms) or in the WT. An MPC can, for example, present a chat interface to the user, to aid them with requirement specification that will be stored in the WT. A different MPC could generate code on behalf of the AC, using specifications from the WT. MPCs are modular -- they are composed of smaller modules to foster reusability of the automation components (e.g., shared interface code). They communicate with the user via an Integrated Development Environment (IDE) plugin, integrating seamlessly with tools that are already familiar to the developer. Modular Pluggable Connectors are further discussed in Section~\ref{sec:arch:mpc}.

The \textit{Autonomy Core} moves one step further in terms of automation capabilities. It can, on its own, assess the situation, plan a series of actions, and carry out the plan, modifying the software and/or the WebAssembly Twin. The AC uses MPCs as its working tools (as a developer would use, e.g., an IDE), while the WT becomes its modeling medium/knowledge base. The decisions in the AC are taken by a mix of neuro-inspired and symbolic AI algorithms -- exploiting the best characteristics of both, in the form of a neuro-symbolic fusion. While machine learning (including neural network models) brings immense intuitive and heuristic capabilities, symbolic reasoning grounds the decisions in the formal knowledge from the WT and guarantees transparency, explainability, and control of the solution. The AC is fully transparent with its decision-making, while its autonomy level is fully controllable by the developer. Human supervision or interaction with the AC is performed via dedicated user interfaces, embedded in the IDE plugin, to make it easily accessible for the developers. The Autonomy Core is further discussed in Section~\ref{sec:arch:ac}.

The shared principle of all these components is their openness and modularity, as the only sustainable path to adoption of such a system is through being open and extensible from the start. This is the same philosophy that is used in, among others, the WebAssembly ecosystem, as discussed in Section~\ref{sec:background:wasm}. Let us now discuss each of the three conceptual components of the proposed architecture in more detail.

\subsection{WebAssembly Twin}\label{sec:arch:wt}

\begin{figure}[htb]
    \centering
    \includegraphics[width=0.8\linewidth]{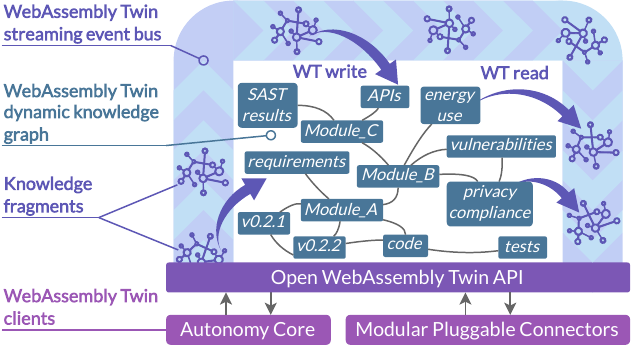}
    \caption{Architecture of the WebAssembly Twin.}
    \label{fig:webassembly-twin}
\end{figure}

The WebAssembly Twin (Figure~\ref{fig:webassembly-twin}) takes a leap beyond the state of the art in software modeling, by aiming to be thoroughly dynamic, modular, and cover any software-related modeling needs. The WT’s content is highly structured and carries semantic meaning, using knowledge graphs~\cite{hogan2021knowledge} as the modeling medium. The meta-knowledge basis for the WT is provided by a set of abstract meta-models, schemas, and ontologies. Where possible, the meta-model will reuse existing, especially standardized ontologies (e.g., from ETSI or W3C), to foster interoperability. Due to its inherent extensibility and the vast collection of already published ontologies, the Resource Description Framework (RDF)~\cite{Cyganiak:14:RCA} ecosystem appears to be most appropriate to be used as the knowledge graph model. In RDF, ontologies are most commonly expressed in the Web Ontology Language (OWL)~\cite{OWL}, allowing one to define expressive reasoning rules. Constraints (useful for, for instance, validation) can be written down in the Shapes Constraint Language (SHACL)~\cite{Knublauch:17:SCL}. RDF, OWL, and SHACL are standardized by the W3C and there are many independent implementations of these standards. This should help in facilitating the adoption of the proposed concept.

This approach opens the way to expressing knowledge about software in a structured manner -- its design requirements, code, tests, interfaces, software contracts, resource use characteristics, security vulnerabilities, and more, on multiple levels of integration. Owing to the intrinsic modularity and flexibility of knowledge graphs, the WT could be easily extended by third parties to cover new use cases, and fuse knowledge across all stages of the software lifecycle. Here, note that the WT cannot be merely a static knowledge graph. It must be a highly dynamic, event-driven system with an internal streaming event bus for carrying commands and knowledge fragments in near-real time. This is needed because the knowledge about the software will be evolving over time. For example, new requirements or target platforms may arise, refactors in the code may be made, and runtime errors may be registered in the deployments. Reacting to and propagating these events quickly is crucial to ensuring that the WebAssembly Twin really does deliver the expected value, instead of being a slow and superfluous add-on to the SE process. In the WT, the event bus writes to and reads from the internal dynamic knowledge graphs and allows for rapid propagation of commands and information to/from WT clients. With a dynamically changing, modular structure, the WT is able to handle a wide variety of challenging scenarios that require fusing complex knowledge across domains.

Dynamicity and complexity are a major technical challenge, with performance being often a serious obstacle in adopting knowledge-rich systems. This should be addressed through the judicious use of state-of-the-art technology, such as low-latency messaging (e.g., NATS\footnote{\url{https://nats.io/}}, Eclipse Zenoh\footnote{\url{https://zenoh.io/}}), high-performance semantic streaming (e.g., the Jelly protocol\footnote{\url{https://w3id.org/jelly}}~\cite{sowinski2022efficient}), and knowledge graph federation. Access to the WT is to be provided via a straightforward open application programming interface (API), enabling third parties to easily implement services that use the WT (e.g., Modular Pluggable Connectors). The WT should be considered a companion to its corresponding WebAssembly module/component. Thus, it should be at least partially distributed with the built application, for example through the currently-developed warg registry protocol\footnote{\url{https://warg.io/}}. These semantic descriptions of software modules could be used by downstream developers, who could then more easily build their own WebAssembly Twins by composing them from already available WTs.

An initial assessment of ontologies that could potentially be used as part of the WT meta-model, points to: SBSON (dependency management)~\cite{eghan2019dependency}, ontologies developed in the MOST project (Software Engineering)~\cite{parreiras2012semantic}, the Software Description Ontology~\cite{garijo2021software}, or ontologies developed in the Software Process Deployment and Evaluation Framework project~\cite{ruiz2015framework}. Not only strictly software-related ontologies are relevant here, though. Ontologies dedicated to, for example, tracking energy use, environmental footprint, work management, and more could be integrated into the WebAssembly Twin, depending on the specific use cases. In the guiding example, for example, it would be useful to include an ontology of workplace safety such as OSHDO-Core~\cite{lawrynowicz2016towards}, to represent regulatory and client requirements. Moreover, as noted earlier, standardized ontologies related to IoT and sensor networks, such as SOSA/SSN~\cite{Phuoc:17:SSN} or SAREF~\cite{garcia2023etsi} are likely to be used.

\subsection{Modular Pluggable Connectors}\label{sec:arch:mpc}

\begin{figure}[htb]
    \centering
    \includegraphics[width=0.8\linewidth]{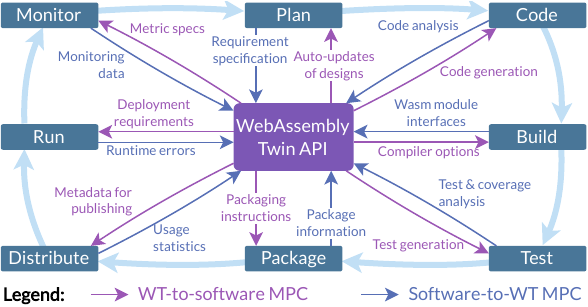}
    \caption{Example Modular Pluggable Connectors.}
    \label{fig:mpc}
\end{figure}

Current software engineering practices involve the use of many diverse tools and processes, such as IDEs, CI/CD systems, code repositories, design tools, and methodologies (DevOps/DevSecOps). These solutions are of tremendous, proven value and must be exploited effectively in the proposed approach. This is done with the Modular Pluggable Connectors -- a set of composable functionality blocks. MPCs, built upon a shared open specification and API, flexibly cover the needs of modifying and reading the WT, interfacing with the aforementioned existing SE tools, and interacting with the user (see Figure~\ref{fig:mpc} for more specific examples). 

The modular structure of MPCs simplifies their implementation -- for example, the code to interact with the WT API can be shared between many connectors, while a pre-trained AI module can be reused for many different tasks. This flexibility enables one to use any kind of AI applicable to a given problem and maximize reuse of existing AI solutions, easing the adoption. This may include: self-supervised generative models~\cite{hou2023large} (e.g., for code and test generation), multimodal deep neural networks~\cite{pawlowski2023effective} (e.g., for interacting with users and reading graphical media such as hand-drawn diagrams), graph neural networks (for WT comprehension, e.g., ULTRA~\cite{galkin2023towards}), time series analysis and prediction models (for example, for anomaly detection in monitoring data~\cite{liu2023gradient}), and symbolic reasoning~\cite{chen2020review} for inferring new knowledge based on what is already stored in the WebAssembly Twin. The MPCs are expected to form an open ecosystem of modules developed by multiple industry and research actors.

In the guiding scenario, the software company would ideally be able to reuse many MPCs published openly by other actors. Additionally, they could relatively easily prepare their own MPCs for integrations with the tools already being used by the development teams. This would allow the company to retain most of its original tools and processes, lowering the cost of adopting the new technology. Use case-specific modules could also be prepared, for example an interface that would explain the software's algorithm using occupational health and safety terminology, by exploiting the cross-domain knowledge stored in the WebAssembly Twin.

\subsection{Autonomy Core}\label{sec:arch:ac}
\begin{figure}[htb]
    \centering
    \includegraphics[width=0.7\linewidth]{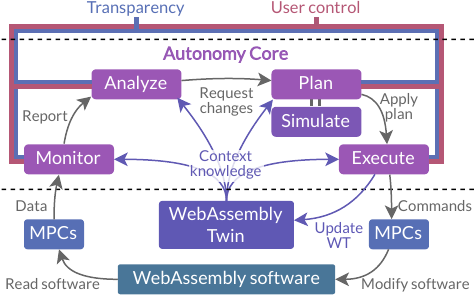}
    \caption{Architecture of the Autonomy Core.}
    \label{fig:autonomy}
\end{figure}

Finally, the leap forward in automation capabilities is provided by the Autonomy Core (AC) (Figure~\ref{fig:autonomy}) -- a general autonomy enabler, capable of planning and executing complex WebAssembly software workflows on its own. The vision behind the AC is to realize complex, self-adaptive use cases -- for example: \textbf{(i)} responding to a spike of runtime errors and fixing the bug autonomously, while also adding tests to prevent the problem from re-occurring; \textbf{(ii)} detecting a security vulnerability in the dependency chain and resolving the issue by updating the affected module; \textbf{(iii)} identifying a possible resource optimization (e.g., more efficient library, unnecessary code) and introducing it in practice.

The Autonomy Core extends the framework of monitor-analyze-plan-execute with knowledge (MAPE-K)~\cite{kephart2003vision}, adapted to fit in the proposed architecture. The WebAssembly Twin serves as the source of knowledge, with the AC also being able to modify the WT in a feedback loop, enabling iterative processes (i.e. iterative self-adaptation). Knowledge graphs were previously proposed as an ideal base for future autonomous systems due to their unique characteristics~\cite{calbimonte2023autonomy}. Using the semantics imbued in the WT, the AC is able to ``understand'' the content of the WT, and generalize it to entirely new concepts (e.g., new use cases, or modeling domains) -- this can be achieved with symbolic reasoning. On the other hand, machine learning methods excel at less rigorous cognitive processes, such as reasoning by analogy. The AC will effectively leverage both, using the emerging neuro-symbolic techniques that aim to bring together the best characteristics of both types of AI~\cite{hitzler2022neuro,lu2024surveying,yu2023survey}.

Maintaining control and accountability of the autonomous processes is crucial, with failures in the AC having the potential for serious social, environmental, and economic consequences. Therefore, a systematic, risk-based system for determining autonomy levels must be employed, considering the potential for errors, and their possible implications. Based on this, the system will use one of the following modes: human-in-the-loop, human-on-the-loop (observing), and full autonomy. In this context, using reinforcement learning with human feedback~\cite{abel2017agent} should be investigated to let the AC dynamically learn from interacting with the user. Transparency and explainability are similarly crucial. Therefore, the AC must be end-to-end transparent and interpretable, with its explanations being based on the well-defined terms in the WT (the symbolic part of the neuro-symbolic hybrid). The WT thus establishes a ``bridge of understanding'' between the user and the AC, providing a set of concepts understandable by both the human and the machine, with the explanation being profiled to suit the nature of the process and the involved stakeholders. This concept of using knowledge graphs to help with machine learning explainability has already been explored in recent studies~\cite{tiddi2022knowledge}.

In the guiding scenario, an autonomous workflow could be used to automatically adapt the software to API changes in the customers' systems. Once such a change was detected, the AC would propose a modification to the integration code and submit it for human review. Alternatively, the AC could monitor the performance of the application in deployment and seek to identify hardware platforms on which the software performs poorly. Subsequently, it could issue improvement recommendations to the development team, along with potential code that could address the issue. The AC could simulate the outcome of different changes to the code and automatically discard modifications that would not improve the performance.

\section{Conclusion and Future Work}

The proposed system architecture aims to tackle several interconnected and very relevant challenges in modern software engineering. By analyzing the current issues concerning multi-architecture development, SE process acceleration, and software modeling, it becomes clear that there is a lot of room for novel methodological concepts and technical innovations. Specifically, the proposed architecture is intended to greatly accelerate software engineering processes, throughout the entire software lifecycle of multi-architecture, continuum-native applications. This is achieved through the proposed combination of dynamic, knowledge-based software models (WebAssembly Twin); flexible and powerful, yet transparent and controllable autonomy (Autonomy Core); and modularized automation components (Modular Pluggable Connectors). The ideas presented here are illustrated in a realistic guiding scenario, originating from the ASSIST-IoT project. 

Realizing the proposed architecture in practice would require further technical and research work, some of which is already happening, as outlined in Section~\ref{sec:background}. The WebAssembly ecosystem is advancing rapidly, with the Component Model, WASI, wasmCloud, and the warg protocol expected to be the biggest drivers of technical progress in the area. The WebAssembly Twin assumes using knowledge graphs in a very demanding scenario, with a lot of dynamic changes and a very complex graph structure. This certainly requires further research in dynamic knowledge graphs~\cite{polleres2023does}, high-performance streaming knowledge graph systems (stream protocols, reasoners, query engines), and interface standardization to connect these solutions together (e.g., the recent RDF Stream Taxonomy proposal~\cite{sowinski2024rdf}). Establishing a sustainable ecosystem of interoperable ontologies for these knowledge graphs is also challenging (cf. issues with reusing ontologies in practice~\cite{sowinski2022ontology}). However, the Semantic Web community has made great progress in implementing the FAIR practices in the recent years~\cite{cox2021ten,garijo2021software,poveda2020coming}, making the outlook rather optimistic. The realization of Modular Pluggable Connectors hinges mainly on defining an open, shared API, and the ability of the community to sustain an ecosystem of interdependent, pluggable modules. Perhaps the most challenging from the research standpoint is the Autonomy Core, which integrates tightly with the WT. Further studies are needed to establish the precise technological and methodological base for such a system, building on the decades of research in autonomous systems. Recently, much work was done in connecting neuro-symbolic AI with knowledge graphs~\cite{delong2024neurosymbolic} and dynamic knowledge graphs~\cite{alam2024neurosymbolic}. Nonetheless, many technical challenges remain to be solved in this field.

\begin{credits}
\subsubsection{\ackname} This study was partially funded by the European Commission, under the Horizon Europe project aerOS, grant number 101069732.

\subsubsection{\discintname}
The authors have no competing interests to declare that are relevant to the content of this article.
\end{credits}

%
\bibliographystyle{splncs04}
\bibliography{bibliography}

\end{document}